\documentclass[aps,prb,twocolumn,groupedaddress,floatfix,nofootinbib]{revtex4-1}

\usepackage{amsmath, mathdots, amsfonts, amssymb, tablefootnote, bm}
\usepackage{todonotes}

\begin{document}
\newcommand{\thermpart}[3]{\left(\!\frac{\partial #1}{\partial #2} \! \right)_{#3}}

\title{Recovering 0 Kelvin Effective Hamiltonian Parameters from High-Temperature Disordered Phases}

\author{Elizabeth Decolvenaere}
\affiliation{Department of Chemical Engineering, University of California Santa Barbara, Santa Barbara, California 93106, USA}

\author{Michael J. Gordon}
\affiliation{Department of Chemical Engineering, University of California Santa Barbara, Santa Barbara, California 93106, USA}

\author{Anton Van der Ven}
\email{avdv@engineering.ucsb.edu}
\affiliation{Materials Department, University of California Santa Barbara, Santa Barbara, California 93106, USA}

\date{\today}

\begin{abstract}
Effective Hamiltonians, when used in tandem with statistical mechanics techniques, offer a rigorous connection between 0 Kelvin ab-initio predictions and finite temperature experimental observations. For alloys, cluster expansion Hamiltonians can coarse-grain out the complex, many-body electron problem of density functional theory, yielding a series of simple site-wise basis functions (e.g., products of site occupancy variables) on an atomic scale. The resulting energy polynomial is computationally inexpensive, and hence suitable for the (tens of) thousands of calculations of large systems required by stochastic methods. We present a new method to run the statical mechanics problem "in reverse", using high-temperature observations and thermodynamic connections to construct an effective Hamiltonian and thereby  predict the 0 Kelvin energy spectrum and associated ground states. By re-examining the cluster expansion formalism through the lens of entropy-maximization approaches, we develop an algorithm to select clusters and determine cluster interactions using only a few, high-temperature experiments on disordered phases. We demonstrate that our approach can recover not only the stable ground states at 0 Kelvin, but also the full phase behavior for three realistic two-dimensional and three-dimensional alloy test-cases. 
\end{abstract}


\maketitle

\section{Introduction}

First-principles electronic structure methods, such as density functional theory (DFT), can provide a unique view into atomic-scale properties that are otherwise \textit{inaccessible} via experiment. Statistical mechanics, or other scale bridging techniques, can then connect the quantum mechanical energy spectrum to the realm of experimentally observable, and industrially-relevant, temperatures and length scales. Directly utilizing first-principles electronic structure methods in statistical mechanics schemes (e.g., to calculate the energy of every microstate), though, is in general computationally intractable. While \textit{ab-initio} molecular dynamics\cite{Pasquarello1998} is increasingly being used to probe high temperature behavior\cite{Ruter2014,Wang2015}, it remains restricted to artificially small periodic unit cells and short simulation times\cite{Mishin2010, Thompson2015}. Instead, atomistic models\cite{Daw1984,Sanchez1984,Tersoff1988,Sanchez1993,Zhong1994,Rabe1995,Zhong1995,Bush1994,Fontaine1994,VanDuin2001,Drautz2004,Zhou2007,Bhattacharya2008,Thomas2013,Wojde2013,Thompson2015,Senftle2016} are more often used to represent a first-principles landscape as a function of relevant degrees of freedom. The path from electronic structure to the laboratory, however, is almost entirely one-way: should an \textit{ab-initio} method prove unreliable when compared to experiment, the experiments cannot be meaningfully used to inform and improve the electronic structure model with the same detail and precision as a direct, first-principles method.

This situation motivates the development of a technique that goes ``in reverse'', whereby measurements of an easily accessible, high-temperature, and disordered phase are used to develop an atomistic model that is accurate at zero Kelvin.  The advantages of such an approach are many. An accurate atomistic model parameterized with high temperature data can be used to predict the energy spectrum over the microstates of a solid, as well as to reveal thermodynamic ground states that are otherwise difficult to determine experimentally (e.g., due to sluggish kinetics at low temperatures)\cite{Zunger2002}. Furthermore, the model can be applied in conventional Monte Carlo simulations to predict the full phase diagram\cite{Burton2012,Puchala2013,Chang2015,Natarajan2016}, or with variance constrained Monte Carlo to predict free energies inaccessible to experiment (e.g., inside the spinodal of a miscibility gap)\cite{Sadigh2012}. Even kinetic properties, such as diffusion\cite{VanDerVen2008, VanderVen2010, Bhattacharya2011} and precipitate nucleation and growth\cite{Vaithyanathan2004}, can be elucidated with such a model.

Effective Hamiltonians\cite{Sanchez1984,Laks1992,Sanchez1993,Zunger1994,Fontaine1994,Tepesch1995,Drautz2004}, which have seen extensive use in the literature\cite{Blum2004,VanderVen2005,Zhou2006,Seko2006,Burton2006,VanDerVen2008,Levy2010,Bhattacharya2011,Burton2012,Puchala2013,Thomas2013,Yuge2014,Belak2015,Chang2015,Chen2015,Decolvenaere2015,Natarajan2016,Goiri2016}, provide a framework well-suited to developing such a model.  They have proven to be powerful tools to extrapolate first-principles energy landscapes and come in many forms.  A harmonic Hamiltonian expressed in terms of inter-atomic force-constants, for example, extrapolates first-principles force-displacement relations to predict phonon properties and vibrational free energies. Cluster expansions\cite{Sanchez1984,Sanchez1993,Fontaine1994} and anharmonic lattice dynamics Hamiltonians\cite{Zhong1994, Zhong1994, Rabe1995, Bhattacharya2008,Thomas2013,Wojde2013} have enabled the first-principles study of alloy phase diagrams and structural phase transitions with Monte Carlo\cite{Zhong1994,Zhong1995,Bhattacharya2008,VanderVen2010,Thomas2014,Natarajan2016,Goiri2016}. In their most rigorous form, an effective Hamiltonian can be formulated as a linear expansion in a set of basis functions, expressed in terms of variables that describe particular atomic degrees of freedom. Alloy Hamiltonians, for example, commonly referred to as ``cluster expansions'', are expressed in terms of polynomials of occupation variables associated with clusters of sites (e.g., pairs, triplets etc.) in the crystal\cite{Sanchez1984,Sanchez1993,Fontaine1994}. The resulting polynomial is computationally inexpensive, and thus well-suited for stochastic methods such as Monte Carlo, which require tens of thousands of energy evaluations to calculate accurate thermodynamic properties.

Here, we explore the possibility of developing an ``experiments-first''  effective Hamiltonian, using high temperature experiments to predict zero Kelvin behavior. We present a new method of parametrizing the Hamiltonian using experimental data of the disordered state instead of zero Kelvin quantum mechanical predictions. The approach not only yields a parameterization of the expansion coefficients, but also suggests the most probable truncation of the Hamiltonian. Overall, the method enables the construction of an accurate atomistic model of crystalline materials suitable for a wide variety of stochastic simulation techniques. Our approach provides a new tool to develop full phase diagrams and probe otherwise difficult-to-measure thermodynamic properties, using only a small number of high-temperature observations of a disordered phase.

\section{A Thermodynamic Approach to Cluster Expansion Parameters}

We illustrate our approach of parameterizing an effective Hamiltonian with high temperature experimental data in the context of a binary A-B crystalline alloy. The approach is, nevertheless, general, and can be applied to any effective Hamiltonian constructed as a linear expansion of basis functions that depend on one or more atomic degrees of freedom  (e.g., local magnetic moments, atomic displacements etc.).

    \subsection{Overview of the Cluster Expansion Formalism}
    Each crystal site, $i$, of a binary solid is occupied by one of two components: A or B. We can assign an occupation variable to each site, $\sigma_i=\pm1$ (e.g., A = +1 and B = -1), such that the arrangement of A and B atoms in a crystal of $M$ sites is completely specified by $\overline{\sigma} = \left\{\sigma_1, \sigma_2, \dots, \sigma_M\right\}$. Each configuration has an occupation-dependent energy $E = E(\overline{\sigma})$ that can, in principle, be calculated with a first-principles electronic structure method. Sanchez \textit{et al}\cite{Sanchez1984,Sanchez1993} showed that the configuration dependence of the energy of a crystal can be expanded in terms of an orthogonal basis of cluster functions $\phi_\delta(\overline{\sigma})$ defined as a product of occupation variables belonging to a cluster of sites in the crystal:
    \begin{equation}
        \phi_\delta(\overline{\sigma}) = \prod_{i \in \delta} \sigma_{i},
    \end{equation}
where $\delta$ is a geometric \textit{cluster} of sites. The energy $E(\overline{\sigma})$ can then be written as:

    \begin{equation}
        E(\overline{\sigma}) = \sum_{\delta \in L} \phi_\delta(\overline{\sigma})V_\delta ,
        \label{Ealld}
    \end{equation}
with the constant expansion coefficients, $V_\delta$, capturing the many-body physics of the interactions among the atoms (or molecules, or vacancies) of the crystal. The expansion coefficients are referred to as ``effective cluster interactions'' (ECIs).

    Clusters (of sites) related under space group symmetry operations of the crystal (including translation) will have the same $V_\delta$. Equation~\eqref{Ealld} can be simplified by collecting clusters with identical $V_\delta$ into groups $\Omega_{\alpha}$, where $\alpha$ represents a prototype of a particular orbit of symmetrically equivalent clusters (e.g., all nearest-neighbor pairs). Equation~\eqref{Ealld} then becomes:

    \begin{equation}
        E(\overline{\sigma}) = \sum_{\alpha} \Phi_\alpha(\overline{\sigma})V_\alpha ,
        \label{E}
    \end{equation}
with $\Phi_\alpha(\overline{\sigma}) = \sum_{\delta \in \Omega_\alpha} \phi_\delta(\overline{\sigma})$. The sum in Equation (\ref{E}) is restricted to symmetrically distinct clusters no larger than the volume of the crystal, with no more than $M$ members. The $\Phi_\alpha$, which we will call \textit{extensive cluster functions}, differ from the correlations $\varphi_\alpha(\overline{\sigma})$ conventionally defined in the literature\cite{Sanchez1984}:

    \begin{equation}
        \varphi_\alpha(\overline{\sigma}) = \frac{\sum_{\delta \in \Omega_\alpha} \phi_\delta(\overline{\sigma})}{m_\alpha N_P},
    \end{equation}
where $N_P$ is the number of primitive unit cells in the crystal and $m_\alpha$ is the multiplicity of cluster $\alpha$ per primitive unit cell. $\Phi_\alpha$ and $\varphi_\alpha$ are then related by a factor of $m_\alpha N_P$, such that $\Phi_\alpha$ scales with the size of the crystal. This extensive property will prove useful in applying Legendre transforms to develop thermodynamic potentials for ensembles of fixed extensive cluster functions.

As a last step, it is convenient to express the cluster expansion of the configurational energy as a scalar product between two vectors:  one vector being the collection of extensive cluster functions, $\overline{\Phi}(\overline{\sigma})$, and the other being the corresponding ECIs, $\overline{V}$. Equation~\eqref{E} can then be expressed as a dot product:

\begin{equation}
    E(\overline{\sigma}) = \overline{\Phi}(\overline{\sigma}) \cdot \overline{V}
    \label{EDot}
\end{equation}
Although formally rigorous, a challenge in making the cluster expansion practical is the determination of numerical values for the expansion coefficients $\overline{V}$. The traditional approach is to use DFT or one of its extensions to calculate the energy of a number of configurations, $\overline{E}{(\overline{\sigma})}_\text{DFT}$, and then inverting Equation~\eqref{E} to determine the ECI using one of many schemes\cite{Connolly1983,Ceder1995,Zunger2002,Vandewalle2002a,Hart2005,Mueller2009,Mueller2010,Cockayne2010,Nelson2013,Nelson2013,Nelson2013a}. The cluster expansion, however, extends over all cluster basis functions, which for a binary alloy having a crystal of $M$ sites is equal to $2^{M}$. The expansion must, therefore, be truncated. The choice of clusters to remain in the expansion is an often-studied problem with no simple solution\cite{Kristensen2014}. Previous work, though, suggests the set of clusters to be sparse\cite{Blum2004,VanderVen2005,Burton2006,Seko2006,Zhou2006,VanDerVen2008,Bhattacharya2011,Burton2012,Puchala2013,Yuge2014,Chang2015,Decolvenaere2015,Belak2015,Natarajan2016}, and many techniques have been developed to choose a few basis functions from a large pool of candidates\cite{VandeWalle2002,Hart2005,Hart2005,Cockayne2010}.


    \subsection{Thermodynamic Relationships}
Experiments are unable to provide direct access to the energies of individual microstates $\overline{\sigma}$. Methods for measuring internal energies or enthalpies return only the average over many microstates. Hence, a method based on an inversion of Equation~\eqref{E}, relying on experimental measurements, is unlikely to be found. However, by assigning a thermodynamic interpretation to the expansion coefficients, $\overline{V}$, of a cluster expansion, other expressions can be derived that relate averages of spatial correlations over clusters of sites, which can be measured with a variety of local or reciprocal probes, to the expansion coefficients $\overline{V}$.

   It is convenient to work in the canonical ensemble (constant temperature $T$, number of sites $M$, and alloy concentration $N_A$), which has as partition function $Z$ and free energy $A$:

    \begin{align}
        Z(T, M, N_A) = \sum_{\overline{\sigma}} e^{-\frac{\overline{\Phi}(\overline{\sigma}) \cdot \overline{V}}{k_b T}} \label{Partition} \\
        A(T, M, N_A) = -k_b T \ln\left[Z \right]\, \label{Free-Energy}
    \end{align}
    The sum is restricted to configurations $\overline{\sigma}$ having fixed composition, and  $k_b$ is the Boltzmann constant. Starting with the canonical free energy, we can produce a number of derivatives, some of which have been discussed in previous work\cite{Livet1987}. We highlight a few that are of practical importance here:

    \begin{align}
        \frac{\partial A}{\partial V_{\alpha}} & = \langle \Phi_{\alpha} \rangle \\
        \frac{\partial^2 A}{\partial V_{\alpha} \partial V_{\beta}} = \frac{\partial \langle \Phi_{\alpha} \rangle}{\partial V_{\beta}} & = \frac{\partial \langle \Phi_{\beta} \rangle}{\partial V_{\alpha}}  = -\frac{\text{cov}[\Phi_{\alpha}, \Phi_{\beta}]}{k_b T} \label{hess} \\
        \frac{\partial^2 A}{\partial V_{\alpha} \partial T} = \frac{\partial \langle \Phi_{\alpha} \rangle}{\partial T} & = - \frac{\partial S}{\partial V_{\alpha}} = \frac{\text{cov}[\Phi_{\alpha}, (\overline{V} \cdot \overline{\Phi})]}{k_b T^2} \label{dcdt}\,
    \end{align}
where $\langle y \rangle = \sum_{\overline{\sigma}} y \frac{\exp \left[{\frac{-\overline{V} \cdot \overline{\Phi}(\overline{\sigma})}{k_b T}}\right]}{Z}$ denotes the ensemble average of $y$ and $\text{cov}[y, z] = \langle y z \rangle - \langle y \rangle \langle z \rangle$ denotes the ensemble covariance of $y$ and $z$. $S$ in Equation~\eqref{dcdt} refers to the entropy.

    Equations~\eqref{hess} and~\eqref{dcdt} are response functions, measuring how the ensemble average of an extensive cluster function, $\langle \Phi_{\alpha} \rangle$, responds to a change in either an ECI, $V_{\beta}$, or the temperature. Equation~\eqref{dcdt} is especially useful after expanding the covariance of the products and rearranging slightly:

    \begin{equation}
        k_b T^2 \frac{\partial \overline{\langle \Phi \rangle}}{\partial T} = \text{\textbf{cov}}[\overline{\Phi},\overline{\Phi}] \cdot \overline{V} \label{matrix}
    \end{equation}
    with $\text{\textbf{cov}}[\overline{\Phi}, \overline{\Phi}]$ denoting a matrix, with each element of this matrix, ${\left(\text{\textbf{cov}}[\overline{\Phi}, \overline{\Phi}]\right)}_{\alpha, \beta}$, corresponding to an ensemble averaged covariance between a pair of extensive cluster functions, $\Phi_{\alpha}$ and $\Phi_{\beta}$. The left hand side of Equation~\eqref{matrix} is a column vector of the temperature derivatives of the ensemble averages of the extensive cluster functions $\Phi_{\alpha}$, multiplied by $k_b T^{2}$.

    Equation~\eqref{matrix} is a crucial component of the approach as it provides a connection between a \textit{measurable} set of variables, $\text{\textbf{cov}}[\overline{\Phi}, \overline{\Phi}]$ and $k_b T^2 \frac{\partial \overline{\langle \Phi \rangle}}{\partial T}$, and a desirable (but immeasurable) set of coefficients, $\overline{V}$. Once values have been measured for the temperature dependence of the extensive cluster functions, and for covariances between pairs of extensive cluster functions, it should in principle be possible to invert Equation~\eqref{matrix} to recover the expansion coefficients $\overline{V}$. These expansion coefficients can then be used in standard statistical mechanics approaches to determine ground states and to calculate the \textit{full} phase diagram. Hence, with only a few measurements, information about the entire phase space can be generated.

    \subsection{Entropy-Maximizing Basis Function Selection}

    While Equation~\eqref{matrix} offers the potential to extract the ECI of a cluster expansion from experimental measurements of extensive cluster functions, $\Phi_{\alpha}$, and their covariances, a direct inversion is, in general, infeasible. Experience with first-principles parameterized cluster expansions shows that these Hamiltonians are typically sparse, converging rapidly as the cluster size of a basis function increases, both in spatial extent and number of sites. Even when the clusters are small, their corresponding ECI may be close to zero. Before Equation~\eqref{matrix} can be inverted, it is therefore necessary to devise a method to determine the ``correct'' sparse set of clusters (i.e., non zero elements in $\overline{V}$). Biased regression schemes (such as $l^1$-norm penalization\cite{Society2007}), while attractive for DFT-based cluster expansions\cite{Mueller2009,Nelson2013a,Nelson2013}, perform poorly when elements in the design matrix (i.e., $\text{\textbf{cov}}[\overline{\Phi}, \overline{\Phi}]$) are correlated\cite{Efron2004}. As the columns in our covariance matrix are themselves correlated, we require an external cluster-selection step that is robust to this feature.

To this end, we again rely on a thermodynamic interpretation of the expansion coefficients $\overline{V}$. The entropy-maximization approach of Jaynes\cite{Jaynes1957} (MAXENT) can be employed to develop a simple metric to judge whether a given cluster should be included or excluded in a final regression scheme to extract the non-zero $\overline{V}$ from Equation~\eqref{matrix}. Treating the ECI as thermodynamic variables, we can re-cast the problem in the form of finding a set of parameters, $\overline{V}$, which satisfy:

    \begin{equation}
        \frac{\partial A}{\partial \overline{V}} = \overline{\langle \Phi \rangle}_\text{obs}, \label{d1}
    \end{equation}
where $\overline{\langle \Phi \rangle}_\text{obs}$ is an \textit{observed} value of the extensive cluster functions. This relation yields a microstate distribution that maximizes the ``information entropy'' of the system, given the constraint that $\overline{\langle \Phi \rangle}$, the ensemble average, is equal to $\overline{\langle \Phi \rangle}_\text{obs}$. The Lagrange multipliers in this constrained maximization problem are, conveniently, the ECIs. Our entropy-maximizing solution is then given by the \textit{stationary points} with respect to $\overline{V}$ of the free energy $\Upsilon$:

    \begin{equation}
        \Upsilon(T, M, N_A, \overline{V}, \overline{\langle \Phi \rangle}_\text{obs}) = A(T, M, N_A) - \overline{V} \cdot \overline{\langle \Phi \rangle}_\text{obs}.\label{d2}
    \end{equation}
Finding the stationary points is as simple as solving $\frac{\partial \Upsilon}{\partial \overline{V}} = 0$ (which returns Equation~\eqref{d1}). When the $\overline{\langle \Phi \rangle}_\text{obs}$ are measured in a thermodynamically stable phase, these stationary points are \textit{maxima}, as proven by the sign of the Hessian of $A$ (and, therefore, of $\Upsilon$) in $\overline{V}$:

    \begin{equation}
        \frac{\partial^2 \Upsilon}{\partial \overline{V}^2} = -\frac{\text{\textbf{cov}} \left[\overline{\Phi}, \overline{\Phi}\right]}{k_b T} \leq 0.  \label{dd2}
    \end{equation}
The strict seminegative-definite nature of the Hessian of $\Upsilon$ for thermodynamically stable phases guarantees a single maximum only. This means that any changes in $\overline{V}$ that increase $\Upsilon$ are moving us towards that global maximum --- there are no local maxima upon which to become trapped. Therefore, if we can evaluate how $\Upsilon$ changes when a cluster is included or excluded, we can use the sign of $\Delta \Upsilon$ to determine if that cluster is moving us towards or away from the MAXENT solution.

A difficulty with Equation~\eqref{d2} is that we do not know the free energy $A$ of the phase in which the $\overline{\langle \Phi \rangle}_\text{obs}$ were measured. However, for a disordered solid solution, we can approximate it by performing a Taylor expansion of $A(\overline{V})$ around the non-interacting crystal ($\overline{V}=0$) corresponding to an ideal solution. To first order:

\begin{multline}
    \Upsilon(T, M, N_A, \overline{V}, \overline{\langle \Phi \rangle}_\text{obs}) \approx
    A_0 + \overline{V} \cdot \left(\overline{\langle \Phi \rangle}_0 - \overline{\langle \Phi \rangle}_\text{obs}\right)\,
\end{multline}
where $A_0$ is the ideal solution free energy, and $\frac{\partial A}{\partial \overline{V}}\big|_{\overline{V} = 0} = \overline{\langle \Phi \rangle}_0$ is the vector of ideal-solution extensive cluster functions, which can easily be evaluated, as the sites of any cluster in an ideal solution are uncorrelated by definition. With the Taylor expansion approximation to $\Upsilon$, the criterion $\Delta_\alpha \Upsilon$ as to whether or not a cluster $\alpha$ should be included is then:

        \begin{multline}
            \Delta_{\alpha} \Upsilon \left(T, M, N_A, \overline{\langle \Phi \rangle}_{\text{obs}} \right)  \approx \\ \left(\overline{\langle \Phi \rangle}_0 - \overline{\langle \Phi\rangle}_{\text{obs}}\right) \cdot \left[\overline{V}_{\text{new}}\left(\overline{\langle \Phi \rangle }_{\text{obs}}\right) - \overline{V}_{\text{old}}\left(\overline{\langle \Phi \rangle}_{\text{obs}}\right)\right] \label{dfdv}\,
        \end{multline}
        where $\overline{V}_{\text{new}}\left(\overline{\Phi}_{\text{obs}}\right)$ and $\overline{V}_{\text{old}}\left(\overline{\Phi}_{\text{obs}}\right)$ refer to the values of the ECIs calculated using Equation~\eqref{matrix} with cluster $\alpha$ included and excluded, respectively. By testing each candidate cluster $\alpha$ for $\Delta_{\alpha} \Upsilon > 0$, we can differentiate between relevant clusters with small ECIs, and clusters with 0 ECIs that recover nonzero values due to regression error. This algorithm is described in Appendix~\ref{algo}, and requires only a single pass through the set of all clusters. For reasons of numerical stability, only cluster observations on the same length-scale of any ``selected'' clusters are used for subsequent evaluations of Equation~\eqref{matrix}.

    The uniqueness of the maximum of $\Upsilon$ is only guaranteed where the free energy varies smoothly, i.e., far from a phase boundary. Additionally, as our Taylor expansion is based around the ideal solution, observations should only be drawn from the disordered phase. This is an easy region to access experimentally, and agrees well with the goals outlined at the beginning of this section. Using Equation~\eqref{dfdv}, we can determine the ideal set of clusters to include, and with Equation~\eqref{matrix}, we can solve for their ECIs. These clusters and ECIs are sparse, thermodynamically-consistent, share a one-to-one mapping with the observed extensive cluster functions, and can be found using only a few observations of the high-temperature, disordered phase.

\section{Testing the Hamiltonian inversion approach on simulated data}
We used simulated data sets to test the viability of the methodology developed in Section II to parameterize an effective Hamiltonian to high temperature measurements.
Benchmarking of the approach was performed on three binary systems (A-B alloys) with their configurational energy described by cluster expansion Hamiltonians. This included two systems on a 2D triangular lattice using: (I) only nearest and next-nearest neighbor (NN and NNN) interactions, and (II) six pseudo-random interactions, including three and four-body clusters. System I has been characterized in-depth by Glosli and Plischke\cite{Glosli1983}. We also studied a 3D FCC lattice (III) using clusters and ECIs generated from first-principles to model the Au-Cu system by Z. Lu, \textit{et al}\cite{Lu1991}. For all systems, we report our results using the following dimensionless, reduced units:
    \begin{align*}
        \tau &= \frac{k_b T}{V_{NN}} & m &= \frac{\mu_A - \mu_B}{V_{NN}} & x_A = \frac{N_A}{M} \\
        v_i &= \frac{V_i}{V_{NN}} & e &= \frac{E}{V_{NN}}\,
    \end{align*}
    where $V_{NN}$ is the nearest-neighbor-pair ECI from the \textit{original} cluster expansion, and $E$ and $e$ refer to \textit{any} type of energy, in the absolute and dimensionless units, respectively. An ECI that is strictly zero, i.e., $v_{\alpha} = 0$, is equivalent to cluster function $\phi_{\alpha}$ being excluded from the cluster expansion.

The normalized chemical potential difference $m$ is related to the slope of the alloy free energy as a function of alloy composition $x_{A}$. The reference states for the model cluster expansions were defined such that the energies for pure A and pure B are both equal to zero. With these reference states, very negative values of $m$ correspond to B-rich alloys while very positive values of $m$ correspond to $A$ rich alloys. Equi-composition alloys have intermediate values of $m$ that are centered around zero.

The simulated data was generated with semi-grand canonical Monte Carlo simulations performed using the three model cluster expansions, using the CASM code\cite{Thomas2013,Puchala2013,VanderVen2010,Puchala2016}.
While the methodology developed in Section II relies upon derivatives taken at constant composition, $x_{A}$,  rather than  at constant chemical potential, $m$, switching from the canonical to the semi-grand canonical ensemble requires only minor modifications to the equations and changes none of the analysis~\footnote{Specifically, the number of configurations $C$ is now $C = 2^N$, and the point-term ECI changes from $V_1$ to $V_1' = V_1 + \frac{\mu}{2}$. This transformation is achieved by examining the argument to the exponential in the semigrand canonical ensemble: $\protect{\frac{-\overline{V} \cdot \overline{\Phi}(\overline{\sigma}) + \mu N_A }{k_b T}}$, and noting that, for our choice of basis set, $N_A = \protect\frac{\Phi_1(\overline{\sigma})+N}{2}$. We can then collect $\Phi_1(\overline{\sigma})(V_1 + \protect\frac{\mu}{2}) = \Phi_1(\overline{\sigma}) V'$ and bring the remaining $\protect\frac{\mu N}{2}$ out of the exponential, and indeed out of any outer sum over microstates, entirely as the number of sites $N$ is unchanging.}.
The Monte Carlo simulations were used to calculate ensemble averages of the extensive cluster functions, $\left<\Phi_{\alpha}\right>$, and their covariances, $\text{\textbf{cov}}[\overline{\Phi}, \overline{\Phi}]$, in the disordered solid solution of the model alloys at high temperature. These two quantities represent the ``experimental data'' needed to invert Equation~\eqref{matrix} to determine the ECIs. A $30 \times 30$ periodic supercell of the 2D triangular lattice was used to simulate data for systems I and II, while a $14 \times 14 \times 14$ periodic supercell of the FCC primitive cell was used to generate data for system (III). All measurements were taken from cooling runs at constant dimensionless chemical potential $m$. The number of passes ($N_\text{pass}$), starting dimensionless temperature ($\tau_0$), incremental dimensionless temperature ($\Delta \tau$), and incremental dimensionless chemical potential ($\Delta m$) are given in Table~\eqref{MC}:

For all three model alloys, we found a strong dependence of the recovered ECI on the value of $m$ used to generate the simulated experimental data sets. Data sets collected at chemical potentials that stabilize B-rich alloys or A-rich alloys (i.e., very negative or very positive values of $m$) were less robust, as changes in $\overline{\left<\Phi\right>}$ became small at near-pure compositions. However, in the chemical potential range that stabilizes a more equi-compositional alloy, a more consistent and reliable set of ECIs could be recovered (provided the values of $m$ and $\tau$ were not too close to a phase transition). To compare the robustness of simulations performed at different values of $m$, we employed the following ``consistency score'' figure-of-merit:

    \begin{equation}
        S_m = \frac{2}{\left\|\overline{v}_{m} - \overline{v}_{m + \Delta m} \right\|+ \left\|\overline{v}_{m} - \overline{v}_{m - \Delta m} \right\|}, \label{score}
    \end{equation}
where $\overline{v}_m$ is the vector of (reduced) ECIs evaluated for a simulation performed at chemical potential $m$, and $\Delta m$ is the chemical potential step size used when performing multiple simulations. $S_m$ corresponds to the (reciprocal of the) average Euclidean norm of ECIs evaluated at three chemical potentials $m$ and $m \pm \Delta m$. This consistency score is used to evaluate when the inversion algorithm ceases to provide reliable results, due to divergences or zeros in $\text{\textbf{cov}}\left[\overline{\Phi}, \overline{\Phi}\right]$ at phase boundaries or compositional extremes. The reciprocal form provides easier interpretation of the results, and maps the ``steadiest'' solutions into the largest scores.

In the following three sections, we summarize the thermodynamic phase behavior of each model system and describe how the inverted cluster expansions compare to the original cluster expansions used to generate the high temperature data sets. For each alloy system, we determined the final $\overline{v}$ with the following process. High temperature measurements (averages of the extensive cluster functions, $\left<\Phi_{\alpha}\right>$, and their covariances, $\text{\textbf{cov}}[\overline{\Phi}, \overline{\Phi}]$) were calculated for a range of $m$ values. The temperature range was chosen to be both narrow and near (but not at) the highest-temperature phase transition. For each value of $m$, a sparse vector, $\overline{v}_m$ was determined using the algorithm of Appendix~\ref{algo}, based on Equation~\eqref{dfdv}. Next, runs were filtered to only include the range of $m$ values centered on $m=0$ for which $S_m$ remained sufficiently large. Using this \textit{reduced} range, only clusters with ECI (entry in $\overline{v}_m$) that were nonzero more than 50\% of the time were kept, forming the ``selected'' set. Finally, ECIs were calculated for the selected set of clusters at each $m$ in the reduced range, with each $m$ being treated independently. By averaging $\overline{v}_m$ of the selected set of clusters over the reduced range of $m$, a final sparse set of ECIs was determined.


\begin{table}[h]
\caption{Simulation conditions for semi-grand canonical Monte Carlo simulations.\label{MC}}
\begin{ruledtabular}
    \begin{tabular}{c c c c c}
        Simulation & $N_\text{pass}$ & $\tau_0$ & $\Delta \tau$ & $\Delta m$ \\ \toprule
        I (2D) & 10,000 & 1.29 & $-2.59 \times 10^{-3}$ & 0.15 \\
        II (2D) & 5,000 & 6.46 & $-12.9 \times 10^{-3}$ & 0.3 \\
        III (3D) & 5,000 & 3.18 & $-3.18 \times 10^{-3}$ & 0.369
\end{tabular}
\end{ruledtabular}
\end{table}

    \subsection{System I:\@ NN and NNN 2D Triangular Lattice}
\begin{figure}[!t]
        \centering
        \includegraphics[width=0.45 \textwidth]{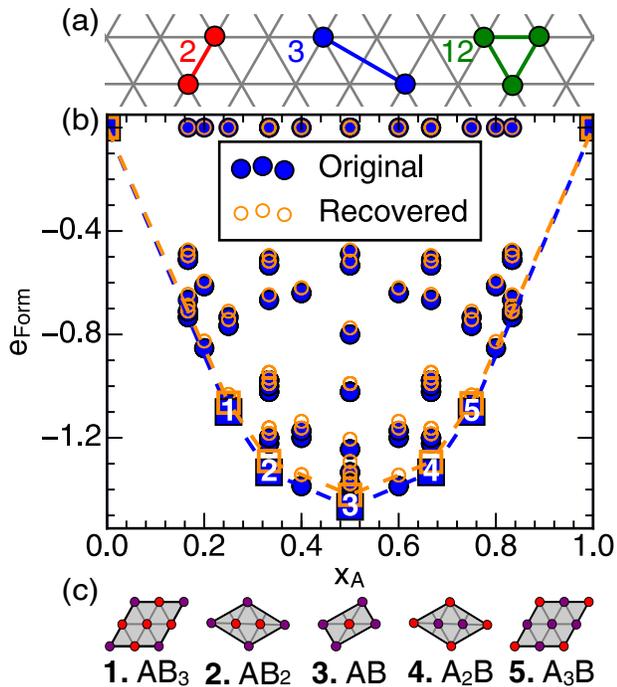}
        \caption{(a) shows the two initial cluster prototypes used in our 2D triangular lattice (2 and 3), in addition to a third recovered cluster prototype (12). (b) shows the composition vs formation energy of a selection of configurations, in supercells containing up to 6 sites. Squares indicate ground states and are numbered to match (c). (c) shows schematic cells of the five ordered ground states. Red circles represent  particle $A$, $\sigma_i = +1$, and purple circles represent particle $B$, $\sigma_i = -1$.\label{Hull_Single}}
    \end{figure}

\begin{figure}[!htb]
        \centering
        \includegraphics[width=0.45 \textwidth]{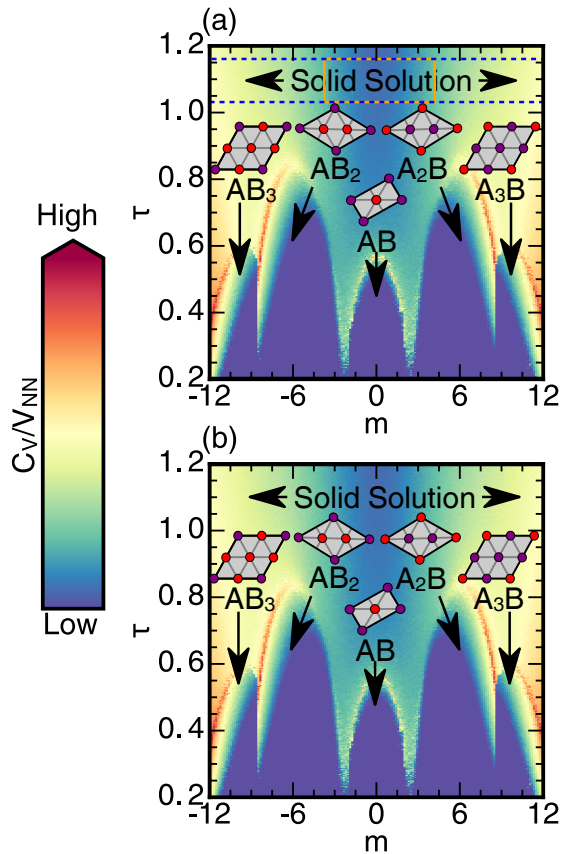}
        \caption{Plots (a) and (b) show logrithmic heatmaps of the heat capacity $C_V$ (scaled by $V_{NN}$), using the original and recovered ECIs, respectively. The approximate phase boundaries visible as sharp shifts in color, and appear at nearly identical locations in both phase maps. The blue dashed lines indicate the range of temperatures across which observations were taken, while the orange lines match those in Figure~\ref{Hist_Single}a.\label{Phase_Single}}
    \end{figure}

    The original and recovered clusters and ECIs for the 2D triangular lattice are given in Table~\ref{Single}, with diagrams of the clusters shown in Figure~\ref{Hull_Single}a. The zero Kelvin formation energies of several structures, including the five ground states for this cluster expansion, are shown in Figure~\ref{Hull_Single}b, with the ordering of each ground state illustrated in Figure~\ref{Hull_Single}c. This set of clusters and ECIs produces a symmetric phase diagram with both first-order and continuous phase transitions as is evident in Figure~\ref{Phase_Single}a. Data for use in our algorithm was sampled over a wide range of chemical potentials and at temperatures from the region in Figure~\ref{Phase_Single}a bounded by blue, dashed lines.

    \begin{table}[!htb]
        \caption{Cluster Characteristics for the 2-Cluster 2D Triangular Lattice.\label{Single}}
        \begin{ruledtabular}
            \begin{tabular}{c c c}
                Cluster $i$ & Original $v_i$ & Recovered $v_i$ \\ \toprule
                2 & 1 & $0.969$ \\
                3 & 0.1 & $0.0966$ \\
                12 & 0 & $1.26 \times 10^{-3}$
            \end{tabular}
        \end{ruledtabular}
    \end{table}

    \begin{figure*}[!htb]
        \centering
        \includegraphics[width=0.95 \textwidth]{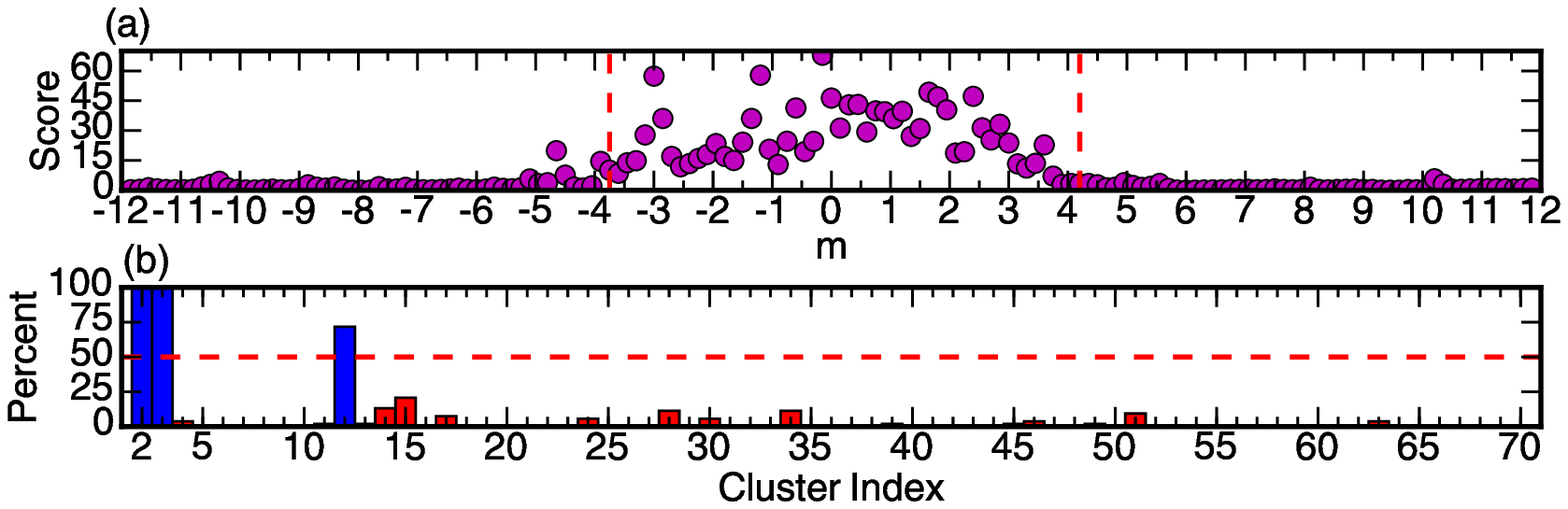}
        \caption{(a) shows the consistency score (Equation~\ref{score}) calculated at each chemical potential (purple dots) using the recovered clusters and ECIs found via our algorithm. Only data between the dashed orange lines was used for subsequent analysis. (b) shows the fraction of runs each cluster appeared in; only clusters above the cutoff ($\geq50\%$, dashed red line) were utilized in the final regression step to determine ECIs.\label{Hist_Single}}
    \end{figure*}

    The algorithm of Appendix~\ref{algo} was applied to data generated over a range of chemical potentials, $m$, yielding a sparse set of ECIs at each $m$. The region where the algorithm performs consistently was determined by the location of the first significant increase in the consistency score, $S_m$, (Equation~\eqref{score}) surrounding $m = 0$, as can be seen in Figure~\ref{Hist_Single}a. The region of data then used to determine the final clusters and their ECIs is indicated by the orange (dashed) lines in Figures~\ref{Hist_Single}a and~\ref{Phase_Single}a, referred to as the ``reliable zone''. The final ECIs were determined following two steps: first, the percentage of $m$ values in the reliable zone in which each cluster was included in the cluster expansion was tallied. This percentage is presented in Figure~\ref{Hist_Single}b. Any clusters which appeared in half or more of the runs in the reliable zone were included in the final set of clusters. This final set of clusters was then used in a global regression over data collected at all chemical potential values $m$ in the reliable zone. The final set of ECI are listed in Table~\ref{Single}.

    In addition to the nearest and next-nearest neighbor clusters (2 and 3, respectively), the algorithm also picked up the nearest-neighbor triplet (cluster 12). The recovered ECIs of clusters 2 and 3 are both within 5\% of their original values, in addition to maintaining the 10:1 ratio present in the original cluster expansion. The nearest-neighbor triplet has a value nearly two orders-of-magnitude smaller than that of the next-nearest neighbor ECI;\@ its impact on any calculated energies is therefore negligible. This assertion is proved by both the zero Kelvin formation energies reproduced using the recovered ECIs in Figure~\ref{Hull_Single}b, and the shape and features of the phase diagram in Figure~\ref{Phase_Single}b. For this simple model cluster expansion, the algorithm of Appendix~\ref{algo} has successfully recovered not only the correct ground states, but the correct phase behavior throughout all of phase space, while utilizing only a tiny fraction of the data available.

    \subsection{System II:\@ 6-Cluster 2D Triangular Lattice}

\begin{figure}[!htb]
    \centering
    \includegraphics[width=0.45 \textwidth]{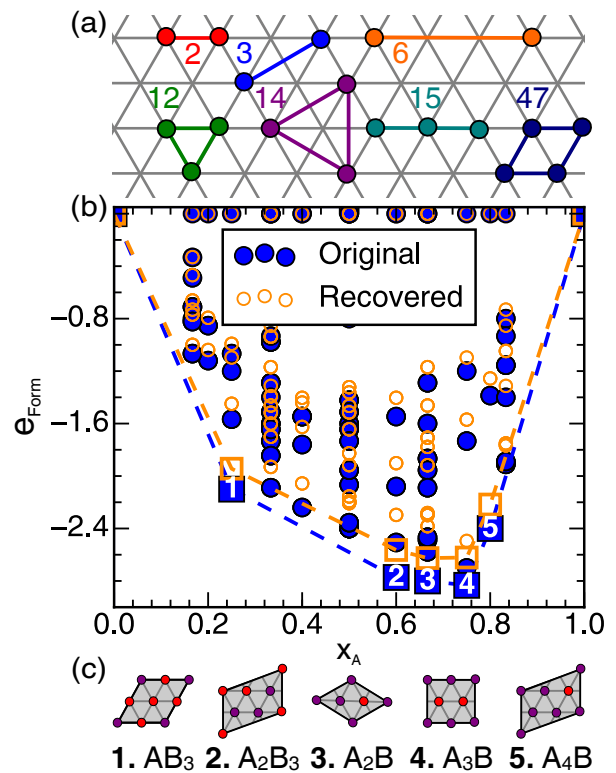}
    \caption{(a) shows the six initial cluster prototypes (2, 3, 6, 12, 15, 47) used to generate data, as well as a spuriously-recovered cluster prototype (14). (b) shows the composition versus formation energy for all configurations in supercells containing up to 6 sites. Squares indicate ground states and are numbered to match (c). (c) shows schematic cells of the five ordered ground states. Red circles represent particle $A$, $\sigma_i = +1$, and purple circles represent particle $B$, $\sigma_i = -1$.\label{Hull_Six}}
    \end{figure}

\begin{figure}[!htb]
    \centering
    \includegraphics[width=0.45 \textwidth]{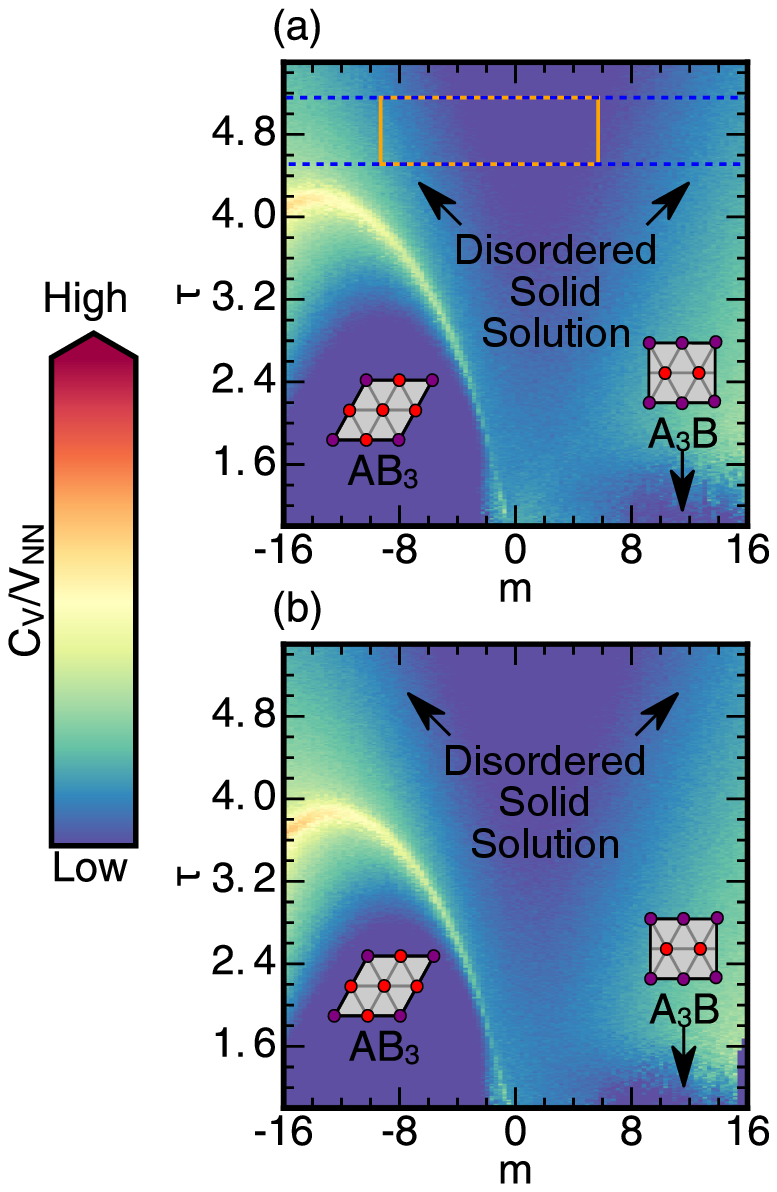}
    \caption{Plots (a) and (b) show logrithmic heatmaps of the heat capacity $C_V$ (scaled by $V_{NN}$), using the original and recovered ECIs, respectively. The approximate phase boundaries are visible as sharp shifts in color, and appear at nearly identical locations in both phase maps, save for a slight amount of scaling. The blue dashed lines indicate the range of temperatures across which observations were taken, while the orange lines match those in Figure~\ref{Hist_Six}a.\label{Phase_Six}}
    \end{figure}

    To examine a more complex cluster expansion for the triangular lattice, six clusters were chosen to represent a spread of cluster lengths and cluster sizes. The values of the ECIs were chosen randomly and are listed in Table~\ref{Six}. Their corresponding clusters are shown in Figure~\ref{Hull_Six}a. Zero Kelvin formation energies for a selection of orderings on the triangular lattice, including the five ground states, are shown in Figure~\ref{Hull_Six}b, with orderings for each ground state illustrated in Figure~\ref{Hull_Six}c. The phase diagram for this cluster expansion is shown in Figure~\ref{Phase_Six}a and is asymmetric, exhibiting both first-order and continuous phase transitions. As before, only data at temperatures bounded by the two blue, dashed lines in Figure~\ref{Phase_Six}a was used in the algorithm of Appendix~\ref{algo} to recover the ECI\@.

    \begin{table}[!htb]
        \caption{Cluster Characteristics for the 6-Cluster 2D Triangular Lattice.\label{Six}}
        \begin{ruledtabular}
            \begin{tabular}{c c c}
                Cluster $i$ & Original $v_i$    & Recovered $v_i$   \\ \toprule
                2           & 1                 & $0.946$           \\
                3           & 0.3               & $0.266$           \\
                6           & 0.5               & $0.433$           \\
                12          & 0.3               & $0.286$           \\
                14          & 0                 & $-0.0109$         \\
                15          & 0.5               & $0.450$           \\
                47          & 0.3               & $0.287$
            \end{tabular}
        \end{ruledtabular}
    \end{table}

    \begin{figure*}[!htb]
    \centering
    \includegraphics[width=0.95 \textwidth]{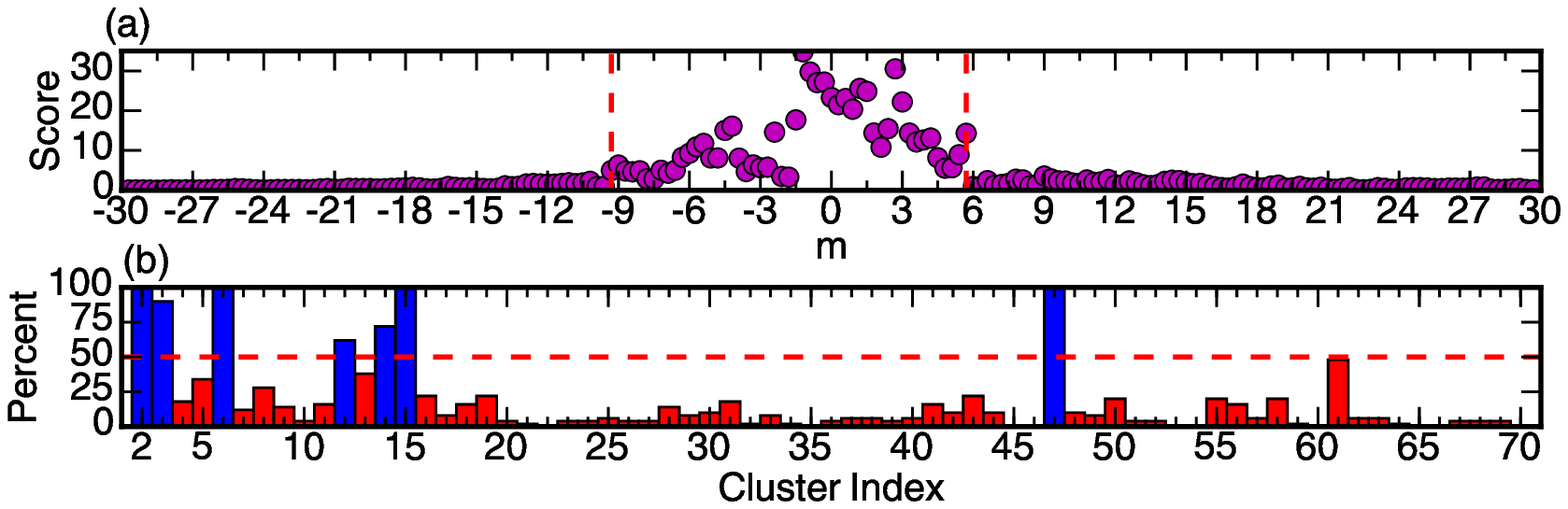}
    \caption{(a) shows the score (Equation~\eqref{score}) calculated at each chemical potential (purple dots) using the recovered clusters and ECIs found via our algorithm. Only data from between the dashed orange lines was used for subsequent analysis. (b) shows the fraction fo runs each cluster appeared in; only clusters above the cutoff ($\geq50\%$, red dashed line) were used in the final regression to determine the ECIs.\label{Hist_Six}}
    \end{figure*}

    Similar to system I described in Section IIIA, we calculated a consistency score for each chemical potential, and used an increase in the consistency score to bound the ``reliable zone''. The scores and resulting boundaries are shown in Figure~\ref{Hist_Six}a. The cluster frequencies in this region are plotted in Figure~\ref{Hist_Six}b, with clusters selected more than 50\% of the time utilized in the final series of regressions. In addition to the original clusters, the next-nearest-neighbor triplet (cluster 14) was picked up, with an ECI one order-of-magnitude smaller then the next-smallest ECI\@. All of the remaining ECIs recovered were within 15\% of their original values, and 10\% of their relative relationships to the nearest-neighbor ECI\@.

    The recovered cluster expansion correctly reproduces the same ground states and formation energies (with a vertical offset) of the original cluster expansion, as shown in Figure~\ref{Hull_Six}b. The calculated phase diagram of Figure~\ref{Phase_Six}b shows that the transition temperatures and the nature of the transition (i.e., first-order versus continuous) are also faithfully reproduced across all of phase space. These results demonstrate the ability of the algorithm to recover a cluster expansion from high temperature data that correctly predicts phase stability over all of phase space.

    \subsection{System III:\@ 3D FCC Lattice}
\begin{figure}[h]
    \centering
    \includegraphics[width=0.45 \textwidth]{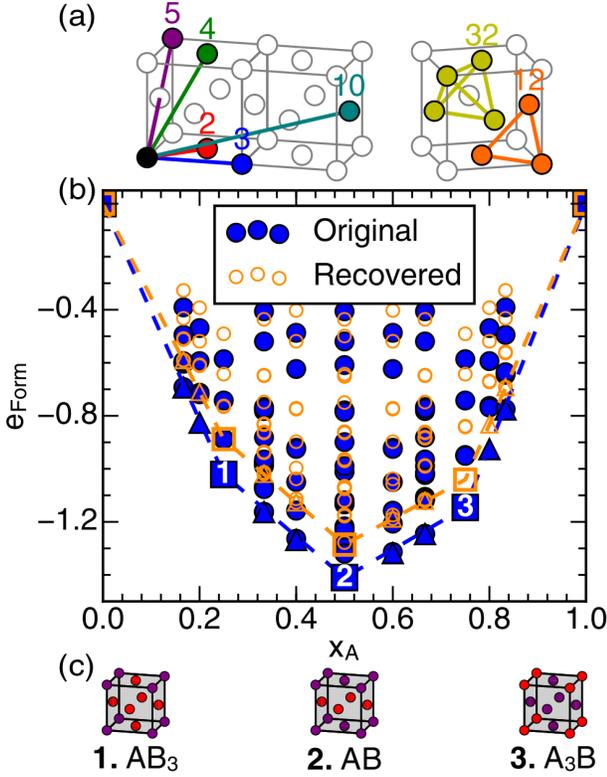}
    \caption{(a) shows the six initial cluster prototypes (2, 3, 4, 5, 12, 32) used to generate data, as well as the newly recovered cluster prototype (10). (b) shows the composition versus formation energy for all configurations in supercells containing up to 6 sites, and selected supercells containing up to 8 sites. Squares indicate ground states and are numbered to match (c), triangles indicate degenerate configurations that lie along, but do not deform, the common tangent between adjacent ground states. (c) shows schematic cells of the three ordered ground states. Red circles represent particle $A$, $\sigma_i = +1$, and purple circles represent particle $B$, $\sigma_i = -1$.\label{Hull_FCC}}
\end{figure}

\begin{figure}[h]
    \centering
    \includegraphics[width=0.45 \textwidth]{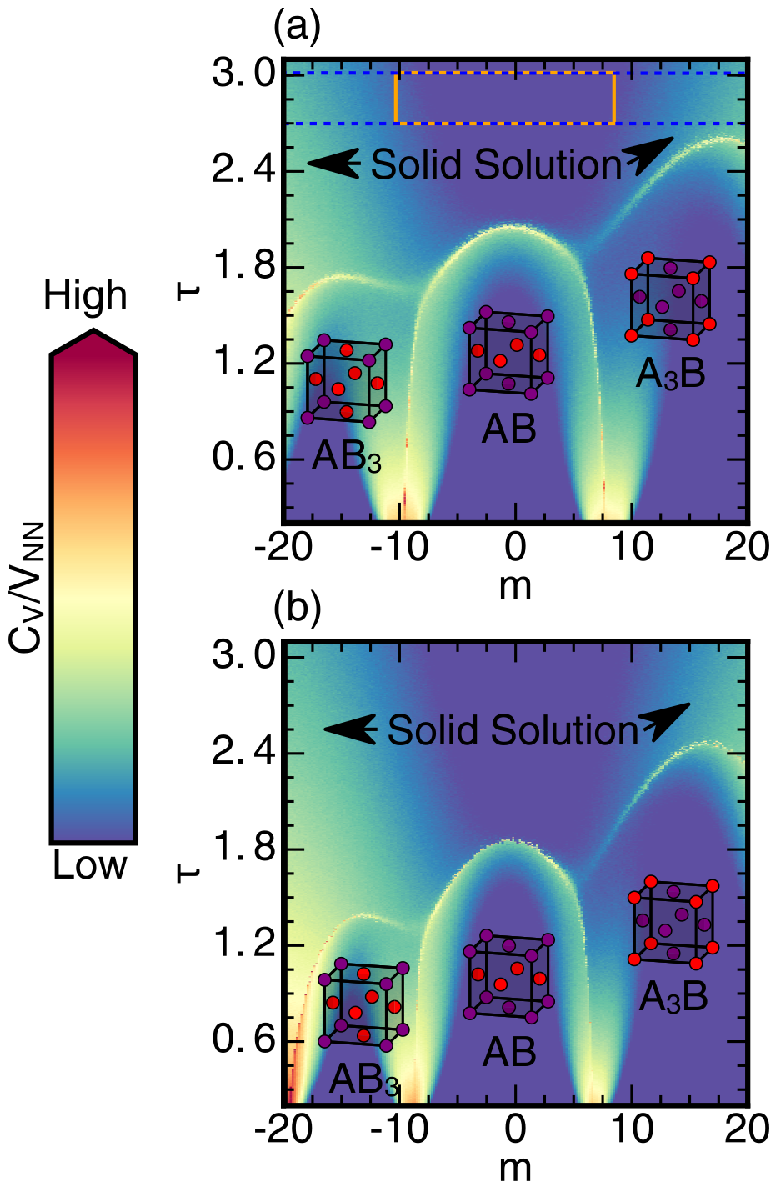}
    \caption{Plots (a) and (b) show logrithmic heatmaps of the heat capacity $C_V$ (scaled by $V_{NN}$), using the original and recovered ECIs, respectively. The approximate phase boundaries are visible as sharp shifts in color, and appear at nearly identical locations in both phase maps, save for a slight amount of scaling. The blue dashed lines indicate the range of temperatures across which observations were taken, while the orange lines match those in Figure~\ref{Hist_FCC}a.\label{Phase_FCC}}
\end{figure}

    The algorithm of Appendix~\ref{algo} was also tested on a cluster expansion constructed by Z. Lu, \textit{et al.}~\cite{Lu1991} to describe the Au-Cu binary alloy. The ECIs are given in Table~\ref{FCC} and the clusters are illustrated in Figure~\ref{Hull_FCC}a. The zero Kelvin formation energies of the L1$_0$ and L1$_2$ ground states, as well as of a number of other configurations, are shown in Figure~\ref{Hull_FCC}b with orderings of the ground states illustrated in Figure~\ref{Hull_FCC}c. Figure~\ref{Phase_FCC}a shows the phase diagram, exhibiting expected behavior akin to that experimentally observed for Au-Cu. As for model systems I and II, only data from temperatures between the two blue, dashed lines in Figure~\ref{Phase_FCC}a was used to recover a cluster expansion.

    \begin{table}[!htb]
        \caption{Cluster Characteristics for the Au-Cu FCC Lattice.\label{FCC}}
        \begin{ruledtabular}
            \begin{tabular}{c c c}
                Cluster $i$ & Original $v_i$    & Recovered $v_i$   \\ \toprule
                2           & 1                 & $0.947$           \\
                10          & 0                 & $0.0170$          \\
                3           & $0.0224$          & $0$               \\
                12          & $0.0622$          & $0.0764$          \\
                4           & $0.0576$          & $0$               \\
                32          & $0.0129$          & $0$               \\
                5           & $0.0157$          & $0$
            \end{tabular}
        \end{ruledtabular}
    \end{table}

    \begin{figure*}[!htb]
    \centering
    \includegraphics[width=0.95 \textwidth]{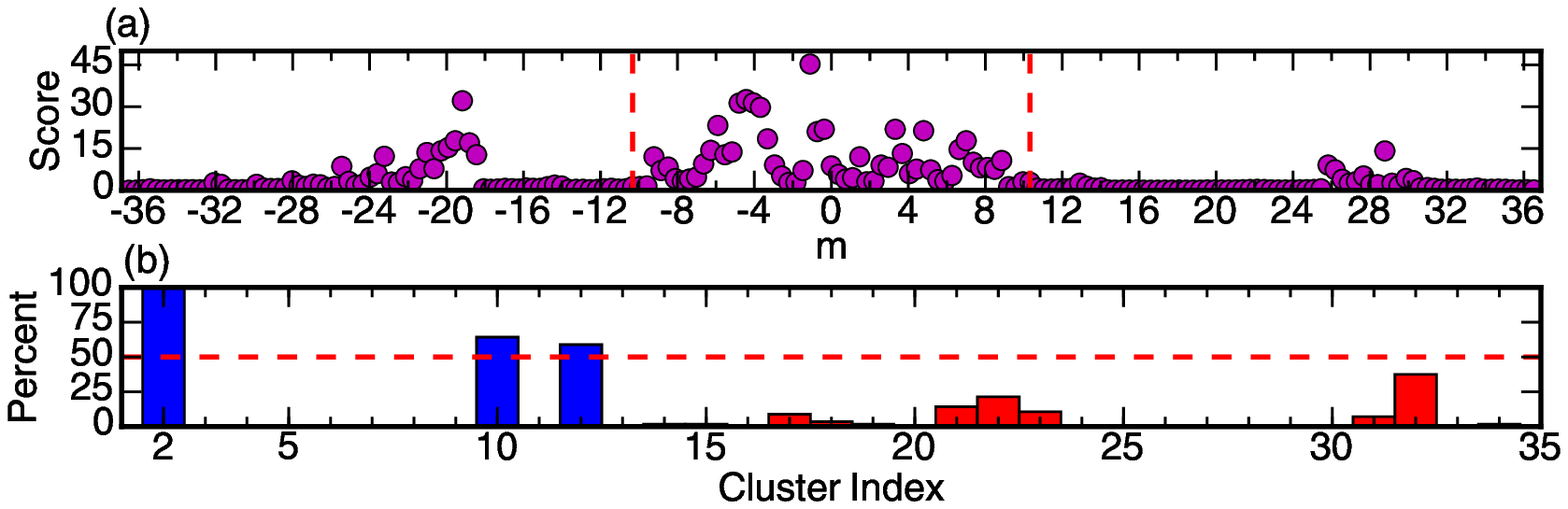}
    \caption{(a) shows the score (Equation~\eqref{score}) calculated at each chemical potential (purple dots) using the recovered clusters and ECIs found via our algorithm. Only data from between the dashed, orange lines was used for subsequent analysis. (b) shows the fraction of runs each cluster appeared in; only clusters above the cutoff ($\geq 50\%$, red dashed line) were used in the final regression to determine the ECIs.\label{Hist_FCC}}
\end{figure*}

The ``reliable zone'' of chemical potentials was again determined using the consistency score (Figure~\ref{Hist_FCC}a) and the final choice of clusters was determined by the frequency with which they were picked up by the algorithm for each chemical potential within this zone (Figure~\ref{Hist_FCC}b). In this case, a \textit{reduced} set of clusters was recovered, excluding the 4-body cluster and many of the 2-body clusters, but including a new longer-range pair interaction, shown as cluster 10 in Figure~\ref{Hull_FCC}a. This new set of clusters correctly reproduces the ground states as well as the on-the-hull degenerate configurations identified using the original cluster expansion as can be seen in Figure~\ref{Hull_FCC}b. Monte Carlo simulations applied to the recovered cluster expansion also faithfully reproduces the phase behavior of the system to within a scaling factor, visible in Figure~\ref{Phase_FCC}b.

While the recovered cluster expansion in this example differs qualitatively from the original one in terms of the number and types of clusters, it nevertheless correctly reproduces the ground states and the finite temperature phase diagram.
Equation~\eqref{dfdv} guarantees a deterministic set of clusters and ECIs, but it does not necessarily guarantee the \textit{same} set of clusters will be picked up as those used to generate the high temperature data. By using the MAXENT method, we bias our recovery towards specific sets of solutions. This example illustrates that multiple sets of clusters and ECIs can generate the same phase behavior. 
Therefore, while the original set of clusters and ECIs can produce the phase diagram in Figure~\ref{Phase_FCC}a, we have recovered another solution with qualitatively the same phase behavior.

\section{Discussion}
We have introduced a method to parameterize an atomistic Hamiltonian that is capable of accurately predicting both the thermodynamic ground states as well as the full phase diagram at finite temperature using only information about the disordered state. We demonstrated the approach for binary alloys modelled with cluster expansion Hamiltonians, which express the dependence of the energy of a multi-component crystal as a linear expansion of cluster basis functions. The foundation of the approach rests on a thermodynamic interpretation of the cluster expansion formalism: extensive cluster basis functions, $\Phi_{\alpha}$, and their corresponding effective cluster interaction (ECI) coefficients, $V_{\alpha}$, form conjugate pairs like any other set of thermodynamic variables. Such an interpretation reveals the existence of Maxwell relations that can be converted to a set of equations (Equation~\eqref{matrix}) relating two sets of measurable quantities (i.e., the covariance between pairs of extensive cluster functions and the temperature derivative of the ensemble averages of extensive cluster functions) to the unknown ECI of a cluster expansion.

Cluster expansions parameterized from first principles tend to be sparse, and not require the complete set of basis functions. We have shown that a thermodynamic interpretation of the ECIs also implies a free energy-like function, Equation~\eqref{d2}. This free energy-like function has a maximum which corresponds to a specified set of basis functions to be retained in a truncated expansion consistent with the \textit{observed} averages of the cluster basis functions. This follows from Jaynes' maximum (information) entropy or MAXENT approach.

The two properties described in Equations~\eqref{matrix} and~\eqref{dfdv} emerge from a thermodynamic interpretation of the cluster expansion formalism. These thermodynamic features motivate and support an iterative algorithm for the parameterization of an effective Hamiltonian to high temperature observations. The final step relies on a regression model to invert Equation~\eqref{matrix}. However, since the measured system is exactly determined (one linear relation and one unknown for each cluster basis function of a cluster expansion), direct inversion of Equation~\eqref{matrix} becomes both numerically unstable and computationally intractable as the number of cluster basis functions becomes exceedingly large in the thermodynamic limit. Furthermore, most multi-component solids can be accurately described with a sparse cluster expansion where only a small subset of the ECI are non-zero. Hence, ordinary least squares is not a suitable method for regression, even if only considering the first $n$ rows and first $m < n$ columns of Equation~\eqref{matrix}. Furthermore, we are also prohibited from sparsity-preserving techniques such as LASSO\cite{Society2007} due to the nature of both the regressors (the covariances) and the observed variable (the change in extensive cluster functions with temperature).  
It is in this context that an initial step involving a maximization of the free energy, Equation~\eqref{d2}, using an approximation for the disordered state, Equation~\eqref{dfdv}, can guide the selection of a sparse truncated cluster expansion (i.e., a sparse set of non zero ECI). Iteration between inverting a sparse form of Equation~\eqref{matrix} and maximizing Equation~\eqref{d2} then leads to a Hamiltonian that is consistent with high temperature measurements of the disordered state. As our three examples illustrated, the Hamiltonians parameterized this way are capable of reproducing the ground state orderings as well as the topology of finite temperature phase diagrams with remarkable accuracy.

The approach introduced here differs from conventional inverse Monte Carlo schemes\cite{Livet1987,Gerold1987,Mosegaard1995,Mosegaard2002,Albert2014}, which seek to recover interaction parameters of a Hamiltonian from measures of average cluster functions. While inverse Monte Carlo methods can generate similar Hamiltonians as the approach introduced here, they require a new round of Monte Carlo simulations for each step in the gradient descent towards the ``correct'' ECIs. Furthermore, inverse Monte Carlo methods provide no proscription as to \textit{which} cluster basis functions to query, leading to instability of the solution when numerous spurious cluster functions are considered simultaneously for the case of a sparse ground-truth. The approach of this work, in contrast, does not require iteration with Monte Carlo and relies on an agnostic approach in the selection of relevant cluster basis functions. In fact, the step relying on Equation~\eqref{d2} can also be incorporated in conventional inverse Monte Carlo schemes as a way of cluster function selection.


We have said much about ``experimental observables'' without yet discussing how the $\overline{\langle \Phi \rangle}$ and $\text{\textbf{cov}}[\overline{\Phi}, \overline{\Phi}]$ may actually be obtained. Ultimately, the extensive cluster functions are merely the products of site occupation variables, and so if provided with exact atomic data, one would map each site onto a lattice, assign a spin variable, and be able to directly calculate $\overline{\Phi}$. The averages and covariances of the clusters can then be calculated by sub-dividing a sufficiently large observation into $N$ smaller observations and taking averages and covariances across this collection of observations. This sort of exact atomic information is available via atom probe tomography, which can yield observations with volumes on the order of $10^6$ nm$^3$\cite{Cerezo2007} or $10^8$ unit cells. High-angle annular dark-field imaging (HAADF-STEM) can also provide atomic-scale resolution of a sample as well, and by varying the depth of focus, a 3D image image can be obtained over a comparable volume\cite{Sohlberg2015}. Atom probe and HAADF-STEM do not provide 100\% coverage of the volumes they query, but each provides a sufficient overabundance of information as to allow for some guesses at the unknown zones. Less directly, information on pair correlations can be obtained via techniques such as x-ray and electron diffraction, the covariances of the pair cluster functions using fluctuation microscopy\cite{Treacy2005}, and short-range pair and multi-body terms using nuclear magnetic resonance (NMR) multiple-quantum experiments\cite{Cadars2006}. However, rather than use diffraction or NMR techniques directly, it is likely these could be used to supplement any interpretation of atom probe or HAADF-STEM analysis, providing better guesses at information that may be missing.

While the approach introduced here has been developed in the context of a binary alloy Hamiltonian, it can be used to invert any effective Hamiltonian that is expressed as a linear expansion of basis functions of relevant degrees of freedom. The linear expansion coefficients that measure the weight of a particular basis function in the effective Hamiltonian can again be interpreted as a thermodynamic variable. Equations similar to~\eqref{matrix} and~\eqref{dfdv} can then be derived that, through an iterative procedure, enable the parameterization of interaction coefficients using measurements in a high temperature phase. The types of Hamiltonians that can be analyzed in this manner include multi-component (i.e., ternary, quaternary, etc.) cluster expansions, spin-cluster expansions describing non-collinear magnetic solids~\cite{Drautz2004} and lattice dynamical Hamiltonians in the harmonic approximation and beyond.\cite{Garbulsky1994,Zhong1995,VanderVen2010,Thomas2013,Thomas2014}

\section{Conclusion}

We have developed a new method to recover relevant interaction coefficients of effective Hamiltonians from experimentally measurable qualities. By careful examination and manipulation of the free energy, a simple mathematical relationship between fluctuations of extensive cluster functions and their related interaction coefficients emerges. The numerical instability of this equation is solved by the development of a secondary criterion, based on the principle of maximum entropy as put forth by Jaynes. Using a single pass through the space of basis functions of the Hamiltonian, we recover a unique solution in polynomial time. The method has been tested in multiple \textit{in-silico} experiments, and faithfully reproduced both the original thermodynamic ground states and the full phase diagrams of each of our simulated systems.

\begin{acknowledgments}

Anirudh Natarajan and John Goiri for many helpful discussions and assistance with performing calculations and Dr.\ Paul Weakliem for help with computational facilities. Elizabeth Decolvenaere is supported by the MRSEC Program of the Natural Science Foundation under Award No. DMR-1121053. Professor Van der Ven is grateful for funding provided by the Office of Naval Research under Grant No. N00014-12-1-0013 (project manager William Mullins). Simulations were performed using resources from the Center for Scientific Computing in the CNSI and MRL, funded by NSF MRSEC (DMR-1121053), NSF CNS-0960316, and Hewlett Packard.
\end{acknowledgments}

\bibliography{library.bib}

\appendix
\section{Selection Algorithm\label{algo}}

    \begin{figure}[!h]
        \centering
        \includegraphics[width=0.45 \textwidth]{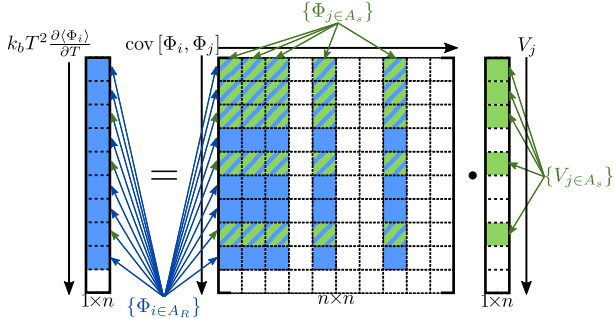}
        \caption{Schematic version of the regression scheme, Equation~\eqref{matrix}, indicating which columns are selected by Equation~\eqref{dfdv} (the set $A_s$, in green) and which rows are included by the cluster radius cutoff (the set $A_R$, in blue). Selecting only rows that represent extensive cluster functions in $A_s$ results in large quantities of data going unused. By using $A_R$, i.e., the set of geometric clusters at or below the cluster radius of the largest geometric cluster in $A_s$, a significantly larger portion of the available observations can be used.\label{rowcol}}
    \end{figure}

    The approach of our algorithm is to invert Equation~\eqref{matrix} via regression, using Equation~\eqref{dfdv} to select which ECIs will be allowed to be nonzero. Additionally, we wish to restrict the range of measurements, i.e., $k_bT\frac{\partial \langle \Phi_i \rangle}{\partial T}$, utilized in our regression. Figure~\ref{rowcol} schematically illustrated the set-up of the regression, indicating which ECIs have been selected (in green) and which measurements are being utilized (in blue). Let some selected set of ECIs be indexed by $A_s = \{\alpha_0, \alpha_1, \dots\}$, such that our model coefficients are $\overline{V}_{A_s} = \{V_{i \in A_s}\}$. Then, let some utilized set of measurements be indexed by $A_R = \{\beta_0, \beta_1, \dots\}$, such that our predicted outputs are $\overline{y}_{A_R} = \{k_b \frac{\partial \langle \Phi_{j \in A_R}\rangle}{\partial T}\}$. Then, the columns utilized in our design matrix, $\bm{X}$, must be $A_s$ and the rows utilized must be $A_R$, such that $\bm{X} = \mathbf{cov}[\{\Phi_{j \in A_R}\}, \{\Phi_{i \in A_s}\}]$. These definitions will be utilized extensively in the description of our algorithm, below. Our selection-and-regression algorithm runs in polynomial time, and requires no additional data or information beyond the same information needed for Equation~\eqref{matrix}. The algorithm as implemented in section III is outlined below:
    

    \begin{enumerate}
        \item Form the set of extensive cluster function indices $A$ for which there exists data, up to a cut-off radius $r$, sorted by increasing geometric cluster\footnote{A \textit{geometric cluster} refers to the set of sites on a lattice included in the definition of a cluster function, but \textit{not} which basis function of the site is being used} size (first by cluster radius, then by the number of sites in the cluster):

            \[A = \left\{\alpha_0, \alpha_1, \dots, \alpha_n\right\}.\]

        \item Initialize the list of selected ECIs $A_s$ by selecting the indices associated with the empty, point and pair clusters from $A$:

            \[A_s = \left\{\alpha_0, \alpha_1, \alpha_2\right\}.\]

        \item Let the cluster radius  $r(\alpha_i)$ be the longest distance between any two sites in $\alpha_i$, and define $R_s$ be the largest cluster radius of geometric clusters represented in $A_s$:

            \[R_s = \max\left[\left\{r(\alpha_i): \forall \alpha_i \in A_s\right\}\right] \]

        \item Form the set of indices affiliated with geometric clusters as small as, or smaller than, $R_s$. This set defines the extensive cluster functions with measurements that we presume to be dominated by signal, rather than noise:

            \[A_R = \left\{\alpha_i: \forall \alpha_i \in A \text{ if } r(\alpha_i) \leq R_s\right\}\]

        \item Using ridge regression~\footnote{\label{ridge} Ridge regression is only used in the \textit{selection} step, not in the for the final ECIs when $A_s$ has been fully determined; the motivation for this choice is described in Appendix~\ref{a_ridge}} (with regularization parameter $\gamma$), calculate the ECIs for $A_s$:

            \[\overline{V}_{A_s} = {(\bm{XX}^\intercal +\gamma \bm{I})}^{-1} \bm{X} \cdot \overline{y}_{A_R} \]

        \item From $A$, select the next index $\alpha_j$ which has not yet been examined, and form the set $A_j$:

            \[A_j = A_s + \left\{\alpha_j\right\}.\]

        \item Using ridge regression\cite{Note2} (with regularization parameter $\gamma$), calculate the ECIs for $A_j$:

            \[\overline{V}_{A_j} = k_b T^2 {(\bm{XX}^\intercal +\gamma \bm{I})}^{-1} \bm{X} \cdot \overline{y}_{A_R} \]

        \item Use Equation~\eqref{dfdv} to calculate $\Delta \Upsilon$:

            \[\Delta_j \Upsilon(T, M, N_A, \overline{\Phi}_{\text{obs}}) = \sum_{\alpha \in A_s} \left(\overline{\Phi}_0^\alpha - \overline{\Phi}_{\text{obs}}^\alpha \right) \left(V_{A_j}^\alpha - V_{A_s}^\alpha \right) \]

        \item If $\Delta_j \Upsilon > 0$, $A_s = A_j$, otherwise $A_s$ remains unchanged.

        \item Return to step 3, until there exist no indices in $A$ which have not been examined.

        \item Using the final set of selected $A_s$ determine the ECIs. Calculate $R_s$ and construct $A_R$ as in steps 3 and 4, respectively, and build $\bm{X} = \mathbf{cov}\left[\left\{\Phi_{j \in A_R}\right\}, \left\{\Phi_{i \in A_s}\right\}\right]$. Solve for $\overline{V}_{A_s}$ using ordinary least squares regression:

            \[\overline{V}_{A_s} = k_b T^2 {(\bm{XX}^\intercal)}^{-1} \bm{X} \cdot \overline{y}_{A_R} \]

    \end{enumerate}

\section{The Effects of Order and Too Much Data\label{order}}
Our selection algorithm provides a unique set of clusters that can describe the observed data; the members of the set do \textit{not} rely on the order in which clusters are considered. As the nature of Equation~\eqref{dd2} does not depend on which subspace of clusters is chosen (the covariance matrix is, regardless, semipositive definite), results for $\Delta \Upsilon_j$ will remain correct in sign even if some clusters have previously been included (or excluded) incorrectly. However, if the measurements (rows) utilized in our regression are dominated by noise, rather than signal, the results of the regression behave erratically. Specifically, we have found that if we include \textit{all} measurements available to us (limited only by when we have chosen to cease enumerating new extensive cluster functions to measure), the results of our algorithm, and in fact, of any regression (even when selecting the set of ECIs used in the underlying Hamiltonian we were trying to recover), were divergent. In this way, the values entering Equation~\ref{dfdv} are then no longer representative of the data, making our scheme (and any scheme) meaningless. Therefore, addition to selecting which ECIs to include in the fit, we have also chosen to restrict which measurements we utilize in our regression; these choices are described in the previous section and illustrated in Figure~\ref{rowcol}.

\section{Motivation for Ridge Regression in Selection\label{a_ridge}}
When performing extensive cluster function selection, we add a penalty term to $A$ proportional to the $l^2$-norm of the ECIs to guarantee numerical stability of both Equation~\eqref{matrix} and~\eqref{dfdv} (i.e., ridge regression). The regularization parameter $\gamma$ is chosen as part of the Bayesian ridge regression scheme as implemented in scikit-learn\cite{Pedregosa2011}. By using ridge regression, we are implicitly assuming that the ECIs are Gaussian distributed with a mean of 0.

    Our second derivatives from Equation~\eqref{dd2} are now:

    \begin{equation}
        \frac{\partial^2 \Upsilon}{\partial \overline{V}^2} = -\frac{\text{\textbf{cov}} \left[\overline{\Phi}, \overline{\Phi}\right]}{k_b T} - \frac{2 \gamma}{k_b T} < 0.  \label{dd3}
    \end{equation}

    For $\gamma > 0$, this guarantees that $A$ (and thus $\Upsilon$) is negative definite (and has a unique maximum). In the regression portion of the selection step, the addition of a regularization term ensures that the modified covariance matrix has no (near-)zero eigenvalues, preserving the numerical stability of the solution. The trade-off, in the form of (uniform) shrinkage, is that the ECIs recovered are slightly smaller than their ``true'' values. Shrinkage of the ECIs provides no penalty in the selection stage as long as the sign of the ECIs is preserved. During the final regression after selection has been performed, only ordinary least squares regression is used to avoid solution bias and shrinkage of the ECIs.
\end{document}